\documentclass[%
reprint,
superscriptaddress,
 amsmath,amssymb,
prl,
]{revtex4-2}

\usepackage{graphicx}
\usepackage{dcolumn}
\usepackage{bm}
\usepackage{hyperref}
\usepackage{xcolor}
\usepackage{float}

\begin{document}

\preprint{APS/123-QED}

\title{Berry phases in  Coulomb drag of double-layer graphene system}

\author{Jianghui Pan}
\affiliation{International Center for Quantum Materials, School of Physics, Peking University, Beijing 100871, China}
\author{Lijun Zhu}
\affiliation{CAS Key Laboratory of Strongly-Coupled Quantum Matter Physics, and Department of Physics, University of Science and Technology of China, Hefei 230026, China}
\affiliation{International Center for Quantum Design of Functional Materials (ICQD), Hefei National Research Center for Physical Sciences at the Microscale, University of Science and Technology of China, Hefei 230026, China}
\author{Xiaoqiang Liu}
\affiliation{International Center for Quantum Design of Functional Materials (ICQD), Hefei National Research Center for Physical Sciences at the Microscale, University of Science and Technology of China, Hefei 230026, China}
\author{Lin Li}
\affiliation{CAS Key Laboratory of Strongly-Coupled Quantum Matter Physics, and Department of Physics, University of Science and Technology of China, Hefei 230026, China}
\affiliation{International Center for Quantum Design of Functional Materials (ICQD), Hefei National Research Center for Physical Sciences at the Microscale, University of Science and Technology of China, Hefei 230026, China}
\affiliation{Hefei National Laboratory, Hefei 230088, China}
\author{Changgan Zeng}
\affiliation{CAS Key Laboratory of Strongly-Coupled Quantum Matter Physics, and Department of Physics, University of Science and Technology of China, Hefei 230026, China}
\affiliation{International Center for Quantum Design of Functional Materials (ICQD), Hefei National Research Center for Physical Sciences at the Microscale, University of Science and Technology of China, Hefei 230026, China}
\affiliation{Hefei National Laboratory, Hefei 230088, China}
\author{Ji Feng}
\email{jfeng11@pku.edu.cn}
\affiliation{International Center for Quantum Materials, School of Physics, Peking University, Beijing 100871, China}
\affiliation{Hefei National Laboratory, Hefei 230088, China}

\date{\today}

\begin{abstract}
  Recent experiments suggest quantum interference effects in the Coulomb drag of double-layer graphene systems. By accounting for correlated interlayer impurity scattering under a weak magnetic field, our theoretical results reveal drag resistivities resembling those in weak (anti-)localization. It is established that the quantum interference effect is most significant when the chemical potentials match. The theory clarifies the roles of intra- and interlayer Berry phases in Coulomb drag in double-layer graphene systems and helps delineate the intra- and intervalley contributions. These insights are valuable for designing graphene-based electronic devices exploiting quantum effects.
\end{abstract}

\maketitle

\textcolor{magenta}{\textit{Introduction.}} 
Coulomb drag, effected by momentum transfer between the charge carriers of two separated conductors\cite{Pogrebinskii77}, offers a non-contact method for characterizing basic properties of a wide range of materials. Coulomb drag measurement is particularly suitable for low-dimensional systems, typically in a double-layer device configuration. Specifically, Coulomb drag has been observed in two graphene layers separated by a dielectric,\citep{CastroNeto09, DasSarma11,Min08,Hwang09, Feldman09, Hirata21,Kim11, Hwang11, Titov13, Song13, Lee16, Liu17, Tse19, Zhu20, Zhu23} as well as in related systems.\citep{Mitra20, Anderson21} In monolayer graphene, quantum interference is also affected by the unusual Berry phase of a Dirac Fermion.\cite{suzuura2002crossover,wu2007weak,lu2015weak} The situation is more complicated in double-layer devices, where the \textit{intralayer} localization, described by the Cooperon propagator, has been argued not to contribute to the drag current.\cite{flensberg1995linear} Recent experimental observations in a graphene double-layer system indicate that the magneto-transresistance deviates from the usual $\sim B^2$ law at low fields, suggesting weak localization (WL)-like or weak anti-localization (WAL)-like behaviors in different doping regimes.\cite{zhu2020frictional, Zhu23} 

Indeed, unlike the intralayer case where momentum conservation ensures that the Berry phase enters into backscattering, such kinematic constraints are apparently absent in interlayer processes. Consequently, the interlayer Berry phase manifested in the interlayer Cooperon \cite{gornyi1999coulomb,garate2012weak} is expected weigh equally into forward and backward scattering, even when correlated scattering of carriers residing on different layers becomes important with a long phase diffusion time $\tau_g$. Meanwhile, it is unclear how the intralayer Berry phase creeps into correlated interlayer scattering processes.

\begin{figure}[h]
  \centering
  \includegraphics[width=\linewidth]{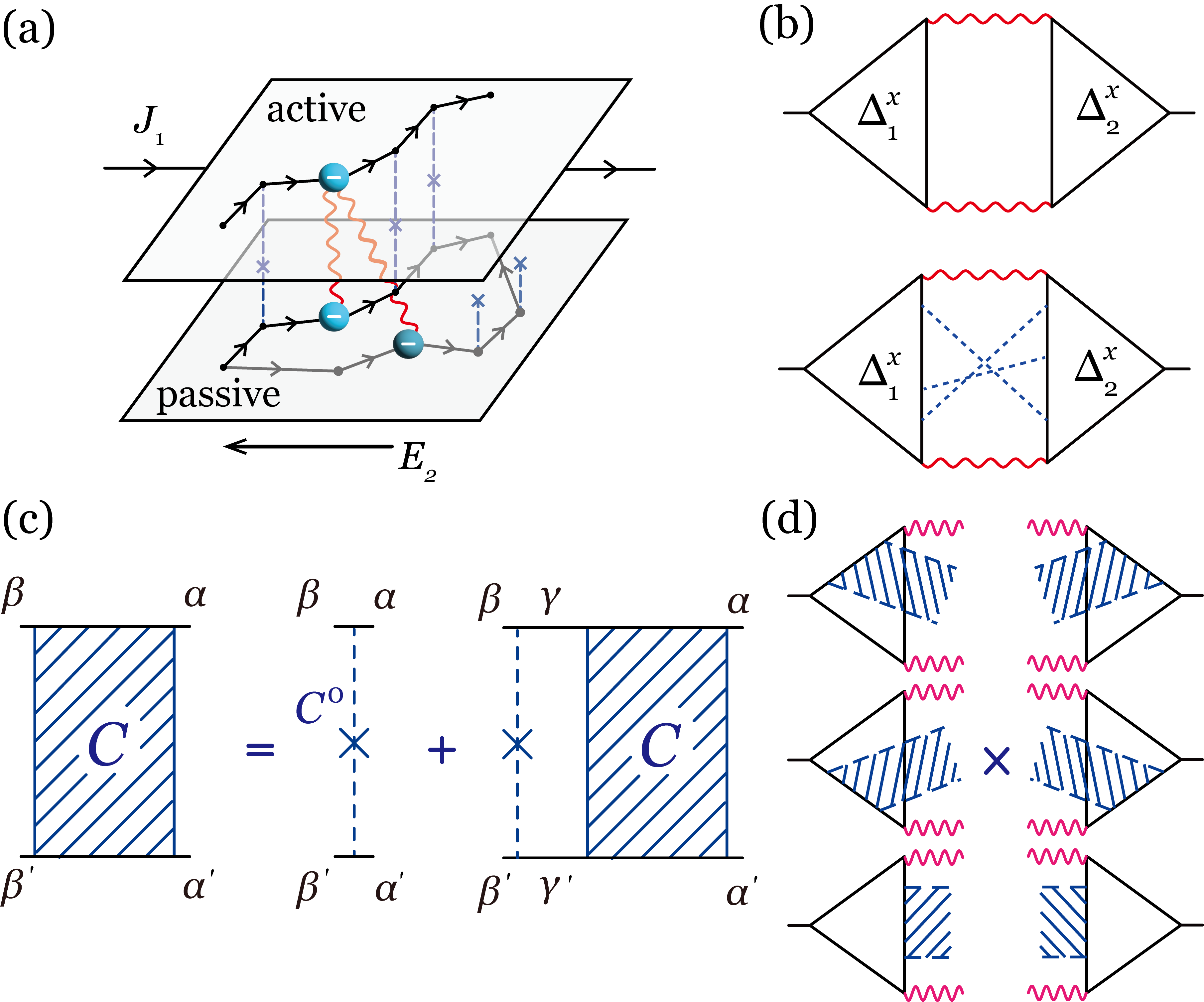}
  \caption{\label{fig:schematic} (a) A schematic diagram of the Coulomb drag effect and the interlayer impurity scattering. (b) Also shown are the Feynman diagrams for the second-order drag conductivity (up) without and  (down) with correlated interlayer impurity scattering. Solid lines represent electron free propagation, while dashed and wavy lines represent the interlayer impurity potentials and the screened Coulomb interaction, respectively. (c) Bethe-Salpeter equation for the interlayer Cooperon. (d) Feynman diagrams for the second-order Coulomb drag with interlayer Cooperons (shaded block).} 
  \end{figure}

In this report, we examine the role of correlated interlayer scattering (CIS) caused by impurities in the Coulomb drag of double-layer graphene systems, considering both intra- and intervalley scattering. Based on finite temperature perturbation theory, we systematically calculate the drag resistivity correction from quantum interference with interlayer Cooperon under a weak magnetic field, and provide the dependence of the correction on chemical potentials, temperature, and interlayer thickness. The microscopic calculation uncovers that there is a remarkable interplay between the \textit{intralayer} and \textit{interlayer} Berry phases in the Cooperon-mediated drag transport, leading to a dominance of forward scattering. Additionally, the correlated scattering time $\tau_g$ for CIS indicates that quantum interference effects are most significant when the chemical potentials match.

\textcolor{magenta}{\textit{Theory.}}
We consider two separated graphene layers coupled solely by Coulomb interaction. The low-energy excitations in a monolayer graphene are described by the usual Dirac Hamiltonian:
\begin{equation}
H_l = v_{\text F}(\xi p_l^x\sigma_x + p_l^y \sigma_y), 
\end{equation}
where $\sigma_i \,(i=x,y,z)$ are the Pauli matrices for lattice pseudospin, $l=1,2$ is the layer index, $\xi=\pm 1$ is the valley index, and $p_l^x$ and $p_l^y$ are the momenta in the absence of interlayer coupling. The eigenvalues of $H_{l}$ are given by  $\varepsilon_{l\alpha }=b_l v_{\text F} k_l$ ($\hbar=1$), with $\alpha$ encapsulating valley $\xi=\pm1$, band $b=\pm1$, and wavevector $\boldsymbol k$. The retarded/advanced single-particle Green's functions read
$
  G^{\text{R/A}}_{l\alpha}(\varepsilon)=1/(\varepsilon-\varepsilon_{l\alpha} \pm\text{i}/2\tau_{l}),
$
where $1/2\tau_{l} = -\operatorname{Im}\Sigma_l$ is assumed to arise primarily from intralayer scattering.

In a double-layer device depicted in Fig. \ref{fig:schematic}(a), the linear current response is generally described by $ J_l^a =\sigma_{ll'}^{ab }E_{l'}^b$,
where $l,l'=1,2$ and $a,b=x,y$. The drag conductivities $\sigma_{12}$ and $\sigma_{21}$ arise from the momentum exchange of charge carriers in the two separate layers via Coulomb scattering. The longitudinal retarded current-current correlation function can be obtained by standard perturbation theory \cite{flensberg1995linear, Kamenev95}, from which the leading-order drag conductivity corresponds to the Feynman diagram in Fig. \ref{fig:schematic}(a). The diagram is second order in the screened Coulomb potential $U(\boldsymbol q)$, and involves two triangular correlation functions $\Delta^a_l(\boldsymbol{q}, \boldsymbol{q}' ; \omega, \omega';\boldsymbol{B})$\citep{flensberg1995linear}, depicted as triangles in Fig. \ref{fig:schematic}(a). Correlated interlayer scattering can be modeled by the interlayer Cooperon inserted between $\Delta_1$ and $\Delta_2$, represented as blue-shaded four-legged vertices in Fig. \ref{fig:schematic}(c), leading to a total 18 diagrams $\{\Pi_i\}$ as shown in Fig. \ref{fig:schematic}(d). The corresponding correction to drag resistivity involving the interlayer Cooperon is then:
\begin{equation}
  \label{eq:delta-sigma}
  \delta \rho_{\text D}=\frac{e^{2}}{2\sigma_{11}\sigma_{22}}\sum_{i=1}^{18}\int \frac{\text d \boldsymbol{q} \text d\omega}{(2\pi)^3} \frac{\partial n_{B}(\omega)}{\partial \omega} \Pi_i(\boldsymbol{q}, \omega;\boldsymbol{B}),
\end{equation} 
where $n_B(\omega)$ is the Bose function and $\boldsymbol{B}$ is the magnetic field perpendicular to the plane.

The interlayer Cooperon differs from the usual intralayer Cooperon. The intralayer Cooperon represents the interference between paths related by time reversal taken by the same electron. Time reversal corresponds to electron backscattering, resulting in constructive interference and localization. As depicted in Fig. \ref{fig:schematic}(a), the interlayer Cooperon describes electrons from different layers moving along superimposed planar paths under the effect of interlayer impurity scattering.\cite{Zhu23} This correlated electron motion can result in a phenomenon similar to weak localization in a single layer. The interlayer Cooperon satisfies the Bethe-Salpeter equation (BSE) depicted in Fig. \ref{fig:schematic}(d). The bare vertex is $C_{\beta\beta'\alpha\alpha'}^0 = \overline{U_{\beta \alpha} U_{\beta^{\prime} \alpha^{\prime}}}$ (overline indicates disorder averaging), evaluated to be $C^0_{\beta\alpha}=\frac{s v^2}{2\tau_i \mu}\left(1+2e^{\text i\phi}+e^{2\text i\phi}\right)$ for intravalley scattering, and $C^0_{\beta\alpha}=\frac{s \xi_\alpha \xi_\beta v^2}{2\tau_i \mu}e^{\text i\phi}$ for intervalley scattering. Here, $\tau_i$ is the interlayer impurity relaxation time, $\phi$ is the angle between $\boldsymbol{ k}_\beta$ and $\boldsymbol{ k}_\alpha$,  and $s=b_\alpha b_\beta=\pm1$.  Due to momentum conservation, we have $\boldsymbol Q = \boldsymbol k_\alpha+\boldsymbol k_{\alpha'} = \boldsymbol k_\beta+\boldsymbol k_{\beta'}$ and $\Xi= \xi_\alpha+\xi_{\alpha'} = \xi_\beta+\xi_{\beta'}$. Since for large $\mu_i$, interband processes are unimportant, the BSE can be written as
\begin{equation}
  \begin{aligned}
  &C_{\beta\alpha}(\boldsymbol Q, \Xi;\boldsymbol{B}) =C^0_{\beta\alpha}(\boldsymbol Q,\Xi)
  +\sum_{\boldsymbol k_\lambda ,\xi_\lambda}C^0_{\beta\gamma}(\boldsymbol Q, \Xi)\\
  &\times G^{\text{ R/A}}_{1,\boldsymbol k_\gamma}(\mu;\boldsymbol{B}) G^{\text{R/A}}_{2,\boldsymbol Q-\boldsymbol k_\gamma}( \mu+\Delta \mu;\boldsymbol{B}) 
  C_{\gamma\alpha}(\boldsymbol Q,\Xi;\boldsymbol{B}).
  \end{aligned}
  \label{eq:bse}
\end{equation} 

In contrast to the single-layer system, the Cooperon depends on the chemical potentials, $\mu_l$, of the two separate layers, which can be tuned independently. The combination of Green's functions that appear on the right-hand side of Eq. (\ref{eq:bse}) depends on $s = \operatorname{sign}(\mu_1\mu_2)$: $G_1^{\text R}G_2^{\text A}$ and $G_1^{\text A}G_2^{\text R}$ for $s=1$, and $G_1^{\text R}G_2^{\text R}$ and $G_1^{\text A}G_2^{\text A}$ for $s=-1$. Under a weak magnetic field, the momentum $\boldsymbol{Q}$ satisfies the semiclassical Einstein-Brillouin-Keller quantization condition $Q_n^2=4eB(n+\gamma/4-\gamma_C/2\pi)$, where $\gamma=2$ is the Maslov index in this system and $\gamma_C=2\pi(1+s)$ is the total intralayer Berry phase. 
Keeping only the divergent part in the weak-scattering limit, the Cooperon has the form:
\begin{equation}
C_{\beta\alpha}(\boldsymbol{Q}) = \frac{sv^2}{\tau\tau_i\mu}\frac{\xi_\alpha\xi_\beta e^{\text i\phi}}{\tau_g^{-1}\pm \mathrm i \Delta\mu+ DQ_n^2}
\label{eq:Cooperon}
\end{equation}
where $\Delta \mu=|\mu_2|-|\mu_1|$ plays the role of a frequency in static limit, $2\mu=|\mu_1|+|\mu_2|$, $2/\tau=1/\tau_1+1/\tau_2$, and the correlated scattering time is
\begin{equation}
\frac{1}{\tau_g}= \frac{\tau_i}{\tau^2}-\left(1-\frac{\Delta \mu}{\mu}\right)\frac{1}{\tau},
\end{equation}
and the interlayer phase diffusion coefficient is $D=v^2\tau_{\text{tr}}/2$. For a divergent Cooperon, it is required that $\tau_i=\tau$  and $\Delta \mu=0$ simultaneously. The angle subtended by $\boldsymbol k_\beta$ and $\boldsymbol k_\alpha$, $\phi$, is the interlayer Berry phase, as it is the geometric phase from CIS, whose role in the drag resistivity will be clarified in the next section.

\textcolor{magenta}{\textit{Role of interlayer Berry phase.}}
Despite the apparent divergence of the Cooperon in the CIS regime, it is not obvious whether or how the Berry phase enters into the drag conductivity. In the monolayer case\cite{suzuura2002crossover}, the conservation of momentum at the current vertices immediately makes the Berry phase of the Cooperon directly relevant to backscattering. In the double-layer drag setup, however, this kinematic constraint is absent. To clarify this issue, we examine the dominant terms of all $\Pi_i$'s, where the Cooperon is inserted between the Coulomb vertices (bottom pair in Fig. \ref{fig:schematic}(d)). As  $Q \rightarrow 0$, the product of the Coulomb vertices 
\begin{equation}
  U(\boldsymbol{q}) U\left(\boldsymbol{q}+\boldsymbol{p}\right) \equiv e^{f(\boldsymbol q, \boldsymbol p)}
  \label{eq:Coulomb2}
\end{equation}
is peaked at $\boldsymbol q_\star=-\boldsymbol p/2$, with $\boldsymbol p =\boldsymbol k_\beta-\boldsymbol k_\alpha$. Applying the Laplace method \cite{butler2007saddlepoint}  to $e^{f}$ in the integral over $\boldsymbol q$ yields
\begin{equation}
  \begin{aligned}
\delta\rho_{\text D} &\approx \frac{- e^2 T^2}{6\sigma_{11}\sigma_{22}}C(\mu)\int\limits_0^{2\pi} \frac{\text d\phi}{2\pi}
\frac{2\pi-\phi-\sin\phi}{\sqrt{\det (-f^{(2)}_\star}) }G_{\boldsymbol k+\boldsymbol{ k}'}^{\text R}G^{\text A}_{\boldsymbol k+\boldsymbol{ k}'}\\
&
\times \cos\phi \left(1+\cos \phi+2|\cos \tfrac{\phi}{2}|\right)^2 \times e^{f(\boldsymbol{q_\star})}
  \end{aligned}
  \label{eq:delta-sigma2}
\end{equation}
where 
\begin{equation}
  C(\mu)= \frac{ v^2}{2D\tau \tau_i \mu} \left[\psi(\frac{\ell^2_B}{2\ell^2_\tau}+\frac{1}{2})-\psi(\frac{\ell^2_B}{2\ell^2_\phi}+\frac{1}{2})\right]
\end{equation}
is for intravalley scattering in which $\ell_\phi$ and $\ell_\tau$ are phase coherence and mean free lengths, respectively. $\ell_B=(2eB)^{-1/2}$ is the magnetic length and $\psi$ is the digamma function. $f^{(2)}_\star$ is the Hessian of $f(\boldsymbol q)$ at $\boldsymbol q_\star$. 

Eq. (7) leads immediately to a simple relation between intra- and intervalley $\delta \rho_D$ when impurity scattering times coincide for ultra-short ranged impurities
  \begin{equation}
    \label{eq:A_relation}
    \eta=\delta\rho_{\mathrm{D},\mathrm{intra}}/ \delta\rho_{\mathrm{D},\mathrm{inter}}=-2.
  \end{equation}
This relation originates from the difference in numbers of intra- and intervalley Cooperons, as can be deduced from the summation in Eq. (\ref{eq:Cooperon}).  In general, it is expected that the inter- and intravalley $\delta \rho_D$ may deviate from this exact relation, but may generally have opposite signs. In what follows, we will discuss exclusively the intravalley $\delta \rho_D$.

The drag resistivity correction exhibits a broadening of approximately  $0.5~\mathrm{T}$ under the magnetic field, displaying a trend similar to that of the intralayer weak localization. The presence of a weak magnetic field primarily influences the overall value of $\delta \rho_D$; therefore, the dependence of $\delta \rho_D$ on various parameters will be investigated without a magnetic field in the following sections.

We have collected terms from the Berry phase on the second line in Eq. (\ref{eq:delta-sigma2}), which reflect the angular dependence of the integrand: while the first term $\cos \phi$ represents the interlayer Berry phase from the Cooperon in Eq. (\ref{eq:Cooperon}), the second term within the parentheses arises from the intralayer Berry phase. Evidently, it is the interplay of both intra- and interlayer Berry phase terms that determines how the CIS impacts the drag resistivity. For calibration, we note that in the monolayer case,\cite{suzuura2002crossover} only $\cos\phi$ enters, but momentum conservation dictates that $\phi\sim\pi$, which leads to WAL/WL in the inter/intravalley processes. In the CIS process in a double-layer system, however, such a kinematic constraint is absent. It is noted that the intralayer term vanishes at $\phi = \pi$, but increases to a maximum at $\phi = 0$. 

It is inferred that the intralayer Berry phase term diminishes backscattering and favors forward scattering. The interlayer Berry phase then determines the overall sign of the drag resistivity, in conjunction with the intralayer Berry phase along with $s$ and valley index. In the weak-scattering regime $\tau\mu\gg 1$ and $B=0$, Eq. (\ref{eq:delta-sigma2}) reduces to
\begin{equation}
\frac{\delta\rho_\text D}{\rho_\text D} \approx -s\frac{48\pi^3}{\zeta(3)}\sqrt{2|\mu|\tau} \log\frac{\ell_\phi}{\ell_\tau}.
\label{eq:delta-simga-2}
\end{equation}

We see that there is a significant CIS correction to the drag resistivity, and $\delta \rho_\text D/\rho_\text D$ depends on temperature only through the temperature dependencies of $\tau$ and $\tau_\phi$.

\begin{figure}[t]
  \centering
  \includegraphics[width=\linewidth]{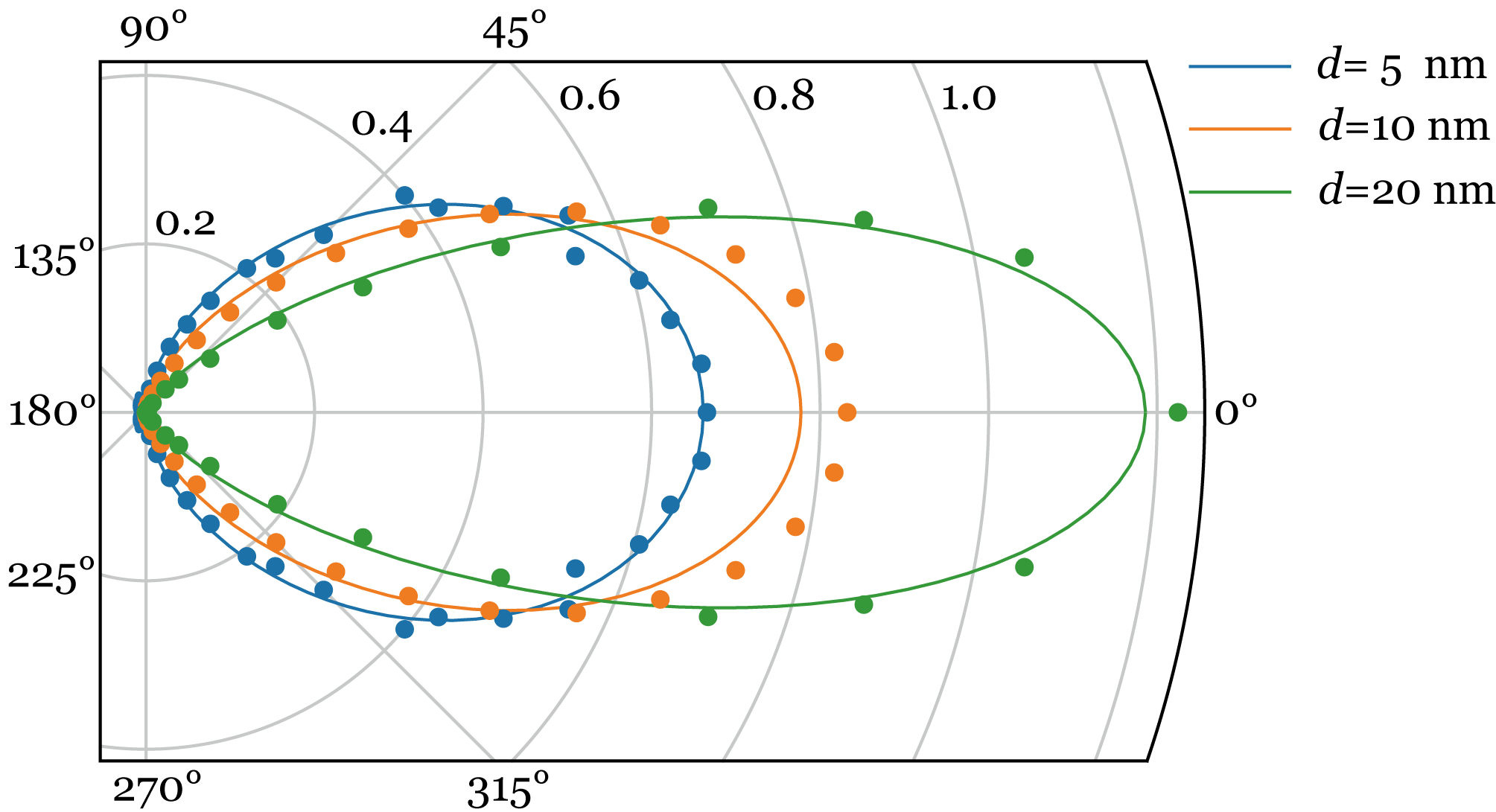}
  \caption{\label{fig:theta} The angular $\phi$-dependence of $\delta \rho_D$ is shown for various $d$ at $T=200\mathrm{K}$ and $A=100$. The integrands of $\delta\rho_D$ are displayed in a polar plot. Dots and lines correspond to the integrands in Eqs. (\ref{eq:delta-sigma}) and (\ref{eq:delta-sigma2}), respectively. The integral values of $\delta\rho_D$ at various $d$ are normalized to 1.} 
  \end{figure}
\textcolor{magenta}{\textit{Numerical results.}} 
While the foregoing analysis reveals the interplay of intra- and interlayer Berry phase terms in the drag resistivity, it is purported to  pertain to the weak scattering regime. To further explore the effects of temperature $T$, interlayer thickness $d$, and impurity strength $A^{-1}$\cite{shon1998quantum} on drag resistivity, we numerically integrate Eq. (\ref{eq:delta-sigma}) using the stratified Monte Carlo technique \cite{lepage2021adaptive}. The interlayer impurity relaxation time is expressed as $1/\tau_i=2\pi\mu/A$, and the intralayer impurity relaxation time $\tau$ is determined through the self-consistent Born approximation\cite{ando2002dynamical}. The screened Coulomb potential takes the form 
\begin{equation}
U(\boldsymbol{q})=\frac{e^2 q}{\epsilon q_{\text s 1}q_{\text s 2}\sinh (qd)}
\end{equation}
in which $q_{\text sl} $ are the Thomas-Fermi wavevectors of layer $l$\cite{tse2007theory}. The summation over $\boldsymbol{Q}$ in the Cooperon will be carried out with the upper and lower momentum cutoffs being $\ell^{-1}_\tau$ and $\ell^{-1}_\phi$, respectively. For simplicity, we estimate that $\ell_\phi$ is four orders of magnitude larger than $\ell_\tau$\cite{DasSarma11}.

Fig. \ref{fig:theta} shows numerical results of the $\phi$-dependence for $\delta \rho_D$, illustrating the impact of the Berry phase and Coulomb interaction. The integrand as a function of $\phi$ in Eq. (\ref{eq:delta-sigma}) is numerically computed (dots in Fig. \ref{fig:theta}), and the integrand of Eq. (\ref{eq:delta-sigma2}) is plotted as lines for comparison. It can be seen that the contribution of forward scattering completely dominates, and strict backscattering is completely suppressed, as expected from our theory. It should also be noted that the Coulomb interaction contributes to the anisotropy by virtue of the Coulomb term in Eq.(\ref{eq:Coulomb2}), which peaks at $\boldsymbol{q}_\star=-\boldsymbol{p}/2$, further favoring forward scattering. This argument is fairly generic, only assuming $U(\boldsymbol{ q})$ is a simple peaked function.

\begin{figure}[t]
  \centering
  \includegraphics[width=\linewidth]{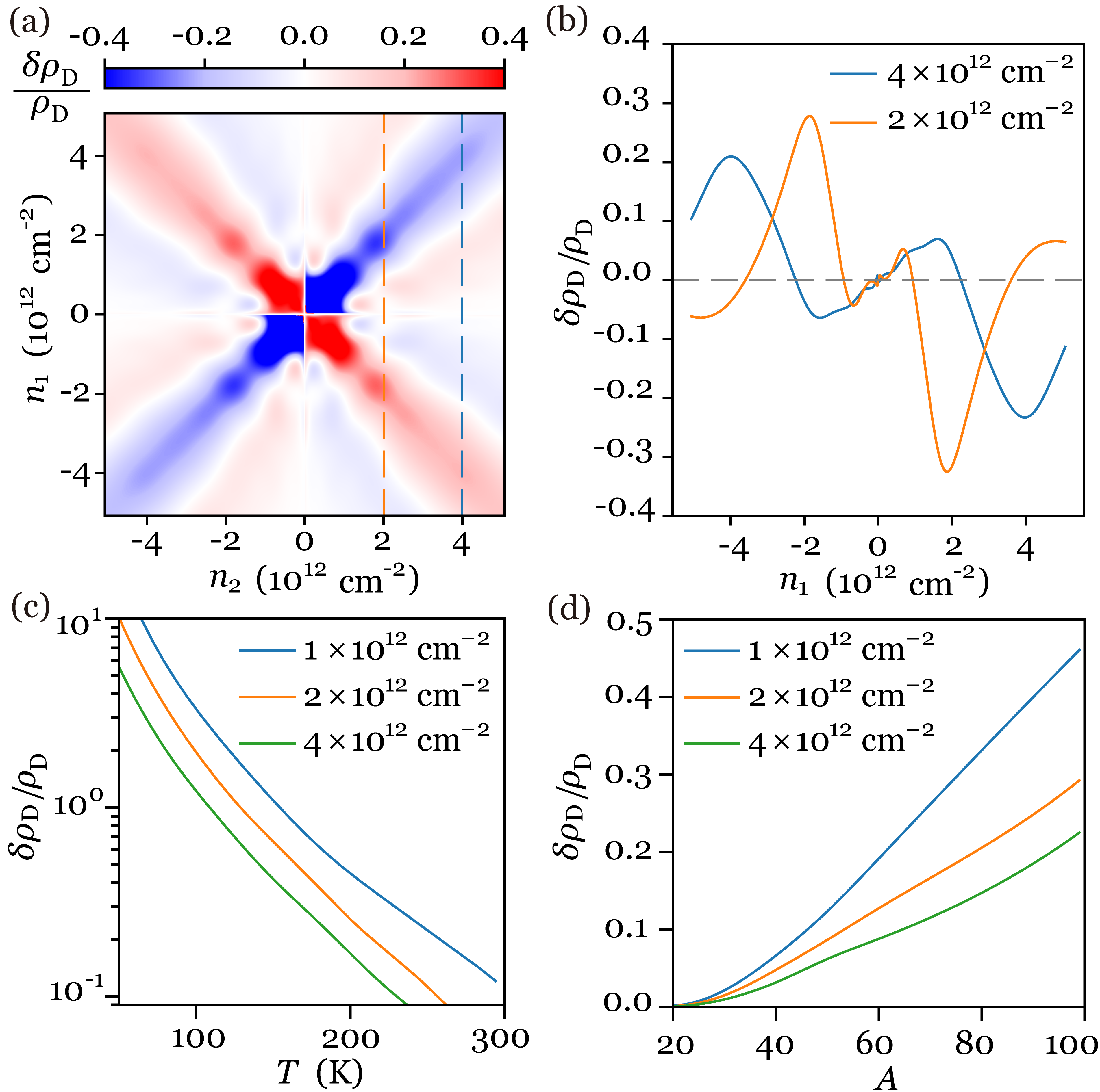}
  \caption{\label{fig:mu}The relationship of $\delta\rho_{\mathrm{D}}/\rho_{\mathrm{D}}$ (a) as a function of $n_1$ and $n_2$ at $T=200~\mathrm{K}, A=100, d=11.3~\mathrm{nm}$, (b) as a function of $n_1$ at various $n_2$. (c) The temperature and (d) the impurity strength dependence of $\delta\rho_{\mathrm{D}}/\rho_{\mathrm{D}}$ for $n=n_1=n_2$ at $A=100$ or $T=200~\mathrm{K}$.}
\end{figure}

Fig. \ref{fig:mu}(a) shows $\delta\rho_{\mathrm{D}}/\rho_{\mathrm{D}}$ as a function of carrier densities in the two layers with only intravalley scattering. Since our model pertains only to intraband impurity scattering. our results are only meaningful at relatively high carrier densities ($|n_l|>5 \times 10^{11}~\mathrm{cm^{-2}}$). We can divide the diagram into four parts based on carrier types (e for $n_l>0$, h for $n_l<0$) in the two layers: e-e, e-h, h-e and h-h sectors. Due to particle-hole symmetry, $\delta\rho_D / \rho_D$ in the four sectors is interrelated by $s=\pm 1$, in accordance with (\ref{eq:delta-simga-2}). The most important feature we find is that the quantum interference mainly occurs where $|n_1|\approx |n_2|$, showing WAL-like behavior in the e-e (h-h) sectors, and WL-like behavior in the e-h (h-e) sectors.

In Fig. \ref{fig:mu}(b), the ratio $\delta \rho_{\text D}/\rho_{\text D}$  is plotted as a function of $n_2$, for two chosen values of $n_1$ (vertical lines in Fig. \ref{fig:mu}(a)). The maximum occurs with identical densities. Note that $|n_l|$ need not be exactly eqaul for significant Cooperon contribution to $\rho_D$, at finite temperatures and small $\tau_g^{-1}$ values. We observe a crossover from WL-like to WAL-like to WL-like again in the e-e (h-h) region, and vice versa in the e-h (h-e) region, although as has been emphasized, the low-density results are not physically relevant. Fig. \ref{fig:mu}(c,d) shows the temperature and impurity strength dependence of $\delta \rho_D / \rho_D$. The key feature is that $\delta\rho_D / \rho_D$ decreases as carrier densities $n$, temperatures $T$ or impurity strength $A^{-1}$ increases. This can be simply understood, as thermal excitation and/or varied impurity strength causes a mismatch in the inter- and intra-layer scattering times, destroying the CIS. 

\begin{figure}[t]
  \centering
  \includegraphics[width=\linewidth]{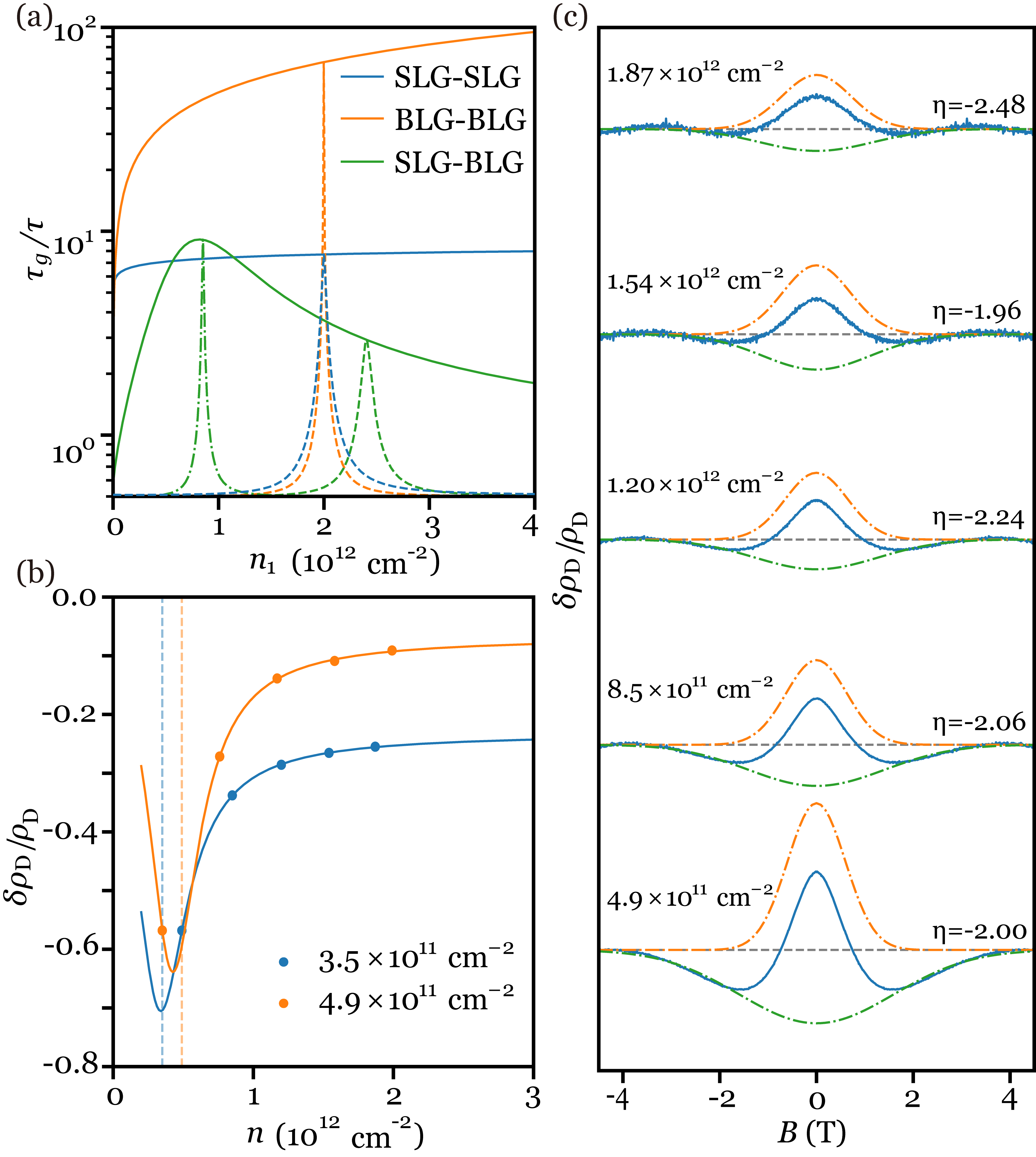}
  \caption{\label{fig:tau_g}(a) $\tau_g$ as a function of $n_1$ in layer-1 with $A=100$. Dashed lines: $\tau_g$ when $n_2=2\times 10^{12} \mathrm{cm^{-2}}$. Solid lines: $\tau_g$ when chemical potentials match. Dash-dot line: $\tau_g$ in SLG-BLG with ${n_2=2.6\times 10^{11} \mathrm{cm^{-2}}}$ which maximizes $\tau_g/\tau$ when Fermi surfaces match. (b) Experimentally measured (dots) $\delta \rho_{\text D}/\rho_{\text D}$ in the BLG-BLG system. The solid lines are Lorentzian fits. (c) Experimental $\delta \rho_D/\rho_D$ with $n_2=0.35 \times 10^{12}~\mathrm{{cm}^{-2}}$ at various $n_1$. The total $\delta \rho_D$ (blue) is partitioned into intra- (orange dash-dot) and intervalley contributions (green dash-dot) (see SM).  }
\end{figure}

It is actually possible to generalize the foregoing discussions on single layer-single layer (SS) systems to other graphene bilayer devices, such as bilayer-bilayer (BB) and single layer-bilayer (SB) systems. The appearance of interlayer quantum interference effects arises from CIS characterized by $\tau_g/\tau$, which is shown for different systems  in Fig (\ref{fig:tau_g})(a). Interlayer coherence occurs when the Fermi surfaces of the two layers coincide. In the case of SS and BB systems, the two layers will have identical carrier densities. In the SB case, the two layers have different carrier densities when the coincidence occurs. As the intralayer Berry phase $2\pi$ also enters into the sign of $\delta\rho_D$ (see Eq. (\ref{eq:delta-sigma2})), the system shows WAL-like behavior in the e-e (h-h) sectors and WL-like behavior in the e-h (h-e) sectors. However, for an SB system, the total intralayer Berry phase is $\pi$, leading to the opposite behavior. 

Finally, we briefly relate the theory to recent experiments \cite{zhu2020frictional, Zhu23}. The experimental $\delta \rho_{\text{D}}/\rho_{\text{D}}$ versus electron density for a BLG-BLG system is shown in Fig. (\ref{fig:tau_g})(b). It is noted that $\delta \rho_{\text{D}}/\rho_{\text{D}}$ increases as the difference in carrier densities diminishes, aligning with theoretical expectations. However, the experimental $\delta \rho_{\text{D}}/\rho_{\text{D}}$ does not vanish at high carrier densities, which cannot be explained by the current theory. Intriguingly, the experimental $\delta \rho_{\text{D}}/\rho_{\text{D}}$ can be perfectly fitted with the sum of two concentric Gaussians with opposite signs, as shown in Fig. (\ref{fig:tau_g})(c). It is found that the ratio $\eta$ of the heights of the two Gaussians falls in the range [-2.5, -1.9]. Relating to Eq. (\ref{eq:A_relation}), we ascribe the two opposing contributions to inter- and intravalley contributions to $\delta \rho$ from correlated interlayer scattering (CIS). It is then inferred from the observation that $\eta \sim -2$ that the experimental devices are closer to the weak scattering limit with strongly localized interlayer impurities.

\begin{acknowledgments}
  We acknowledge the financial support from the National Key R\&D Program of China (Grant No. 2021YFA1400100 and 2023YFA1406300), the National Natural Science Foundation of China (Grants No. 12274003 and 11934001), the Innovation Program for Quantum Science and Technology (Grant No. 2021ZD0302600), and the CAS Project for Young Scientists in Basic Research (Grant No. YSBR-046).
\end{acknowledgments}

\appendix

\bibliography{refs}

\end{document}